\documentclass[prab,reprint,superscriptaddress]{revtex4-2}
\usepackage[switch,columnwise]{lineno}
\usepackage{hyperref}
\usepackage{lipsum}  
\usepackage{amsmath,amssymb}
\usepackage{graphicx}
\usepackage{siunitx}
\usepackage{color,soul}

\newcommand{\etal}{\emph{et al.}}
\renewcommand{\hl}[1]{#1}
\begin{document}

\title{Injection tolerances and self-matching in a quasilinear wakefield accelerator}
\author{J. P. Farmer}
\affiliation{Max Planck Institute for Physics, Munich, Germany}
\affiliation{CERN, Geneva, Switzerland}
\author{L. Liang}
\affiliation{University of Manchester, Manchester, UK}
\author{R. Ramjiawan}
\affiliation{CERN, Geneva, Switzerland}
\author{F. M. Velotti}
\affiliation{CERN, Geneva, Switzerland}
\author{M. Weidl}
\affiliation{Max Planck Institute for Plasma Physics, Garching, Germany}
\author{E. Gschwendtner}
\affiliation{CERN, Geneva, Switzerland}
\author{P. Muggli}
\affiliation{Max Planck Institute for Physics, Munich, Germany}

\begin{abstract}Particle acceleration in a quasilinear plasma wake provides access to high acceleration gradients while avoiding self-trapping of the background electrons.  However, the plasma response to the externally injected witness bunch leads to a variation of the focussing fields along the bunch length, which can lead to a emittance growth.  In order to investigate the impact of this emittance growth on the overall beam quality, we develop a single figure of merit based on a potential high-energy application for the AWAKE experiment at CERN.  We show that the development of such a figure of merit naturally gives rise to constraints on both the tunability and stability of the initial witness bunch parameters.  It is further shown that the unique physics of the quasilinear wake gives rise to broad tolerances for the witness bunch radius at the injection point, as the plasma wakefields self-match to the witness bunch.\end{abstract}

\maketitle

\section{Introduction}
Wakefield acceleration in plasma has attracted considerable attention \cite{lwfa-esarey-review2009,pwfa-muggli-review} since it was first proposed by Tajima and Dawson in 1979 \cite{lwfa-tajimadawson}.  A driver, either a laser pulse or a charged particle beam, excites a plasma wave, which in turn is used to accelerate a trailing witness bunch.  Since plasma is used for the accelerating medium, the acceleration gradients can be orders of magnitude larger than those supported by conventional RF accelerators.

The energy gain which can be achieved in a wakefield accelerator depends on a number of factors.  For the laser-driven case, the accelerated witness bunch will begin to catch up with the driver, with this dephasing setting a maximum distance over which acceleration can occur.  For an electron driver, acceleration is limited by the depletion of the driver energy.  In both cases, these limitations could be overcome by repeatedly re-injecting the accelerated witness bunch into further acceleration stages, each with a fresh driver\cite{lwfa-steinke-staged}.  Alternatively, acceleration to high energies in a single stage could potentially be achieved through the use of a structured driver \cite{pwfa-pukhov-coaxialchannels,lwfa-palastro-flying}.  Of course, both of these options result in an increasingly complex configuration.

The AWAKE experiment at CERN \cite{pwfa-AWAKE-acceleration-cropped} instead makes use of a high-energy proton driver, which essentially removes the constraint of driver depletion \cite{pwfa-caldwell-protondriven}.  However, large acceleration gradients require high plasma densities, which in turn mean high plasma frequencies.  Available proton bunches are therefore too long to effectively drive a plasma wave, with the driver suppressing its own wakefield.  However, the transverse-two-stream instability in plasma \cite{physics-lawson-beams,pwfa-kumar-selfmodulation} causes such long drive beams to ``self-modulate'', i.e.\ break into a train of microbunches \cite{pwfa-lotov-smi}, which can resonantly drive a plasma wave.  The modulated proton beam excites a quasilinear wake \cite{lwfa-schroeder-physics}, below the threshold for wavebreaking and self-injection \cite{plasma-gordienko-wavebreaking}.  This process is similar to the proposed use of self-modulation for laser-driven acceleration \cite{lwfa-andreev-sm,lwfa-krall-sm}, before the widespread availability of ultrashort, ultraintense laser sources.  The self-modulation of the proton beam can be seeded in order to control the phase of the accelerating wakefield \cite{pwfa-batsch-seeded-cropped}.  Further, the nonlinear evolution of the process could be exploited to avoid dephasing between the driver and witness over long acceleration lengths \cite{pwfa-pukhov-selfmodulationphase}.

This quasilinear plasma wake is an attractive scheme for plasma-based acceleration, providing high field gradients while avoiding self-trapping of the background plasma electrons.  Although this regime arises naturally for the modulated proton beam used in AWAKE, this property may also be attractive for a staged laser-wakefield accelerator in order to avoid the generation of dark-currents \cite{physics-rosenzweig-beams}.  However, unlike the blowout regime in which plasma electrons are completely expelled from the region which the witness occupies \cite{lwfa-pukhov-broken}, the local plasma density in the region of the witness bunch will depend on the plasma response to the witness bunch itself.  This results in a variation in the focussing field along the witness bunch length.

In order to avoid betatron oscillations of the witness bunch radius, the transverse size of the witness bunch should be matched to the focussing field which acts upon it.  For the case of a blowout, the transverse gradient of the focussing field is constant, arising due the unshielded ion charge.  The matched radius for the ion channel can then be calculated as:
\begin{align}\label{Eq:match}
  \sigma_\mathrm{ic} = \left(\frac{2c^2{\epsilon_x^\prime}^2}{\gamma\omega_p^2}\right)^{1/4}.
\end{align}
where $\epsilon_x^\prime$ is the normalised emittance and $\gamma$ the Lorentz factor of the witness bunch.  For the quasilinear regime, the matching condition will vary along the bunch length, and, if the witness bunch is sufficiently dense, it may drive its own blowout.  If the bunch is short, the blowout will form behind the witness bunch \cite{pwfa-romeo-assistedmatching}.  In this case, the transverse wakefields do not vary significantly along the bunch length, and a regime of ``assisted matching'' may be accessed, in which betatron oscillations of the witness bunch are sufficiently small that its emittance is conserved.  For a longer witness bunch, the tail may occupy the blowout driven by the head.  In this regime, the radius of the bunch may be matched to the focussing field provided by the ion column, allowing the emittance to be preserved for the fraction of the beam inside the blowout \cite{pwfa-olsen-inject}.  However, betatron oscillations of the head, which experiences a lower focussing field, can result in emittance growth for the leading section of the witness bunch.

This emittance growth, which varies along the length of the witness bunch, is determined by the plasma response to the witness bunch itself.  However, varying the initial witness bunch parameters can also impact on the charge capture, energy gain and energy spread.  In order to investigate how the plasma response can impact on the overall beam quality, we therefore develop an applications-based figure of merit which provides a single metric for the beam quality.  We show how such a metric naturally leads to constraints on both the tunability and stability of the initial parameters of an externally injected witness bunch.

We show that the unique physics of the quasilinear regime leads to large tolerances for the witness bunch radius at the injection point, with the focussing fields ``self-matching'' to the bunch radius over a wide range of initial bunch radii.  This process promises to greatly reduce the technical constraints of realising such a scheme, as the high focussing fields in plasma result in a small matched radius.

The paper is laid out as follows. Section~\ref{simconf} details the simulation model used for these studies, along with some initial simulations.  In section~\ref{fom}, these simulation results are used to motivate a single figure-of-merit for the accelerated witness bunch, which is demonstrated for the case of optimising the witness bunch delay for different charges.  The figure of merit is then used in section \ref{match} to study the impact of the initial witness bunch radius on the acceleration process, illustrating the self-matching of the wakefields.  Additional studies for constraints on the focal position, duration and initial emittance, as well as the influence of the accelerating wakefield amplitude, are shown in section~\ref{everythingelse}, before conclusions are drawn in section~\ref{conc}.

\section{Simulation Configuration}\label{simconf}
\begin{figure}
  \includegraphics{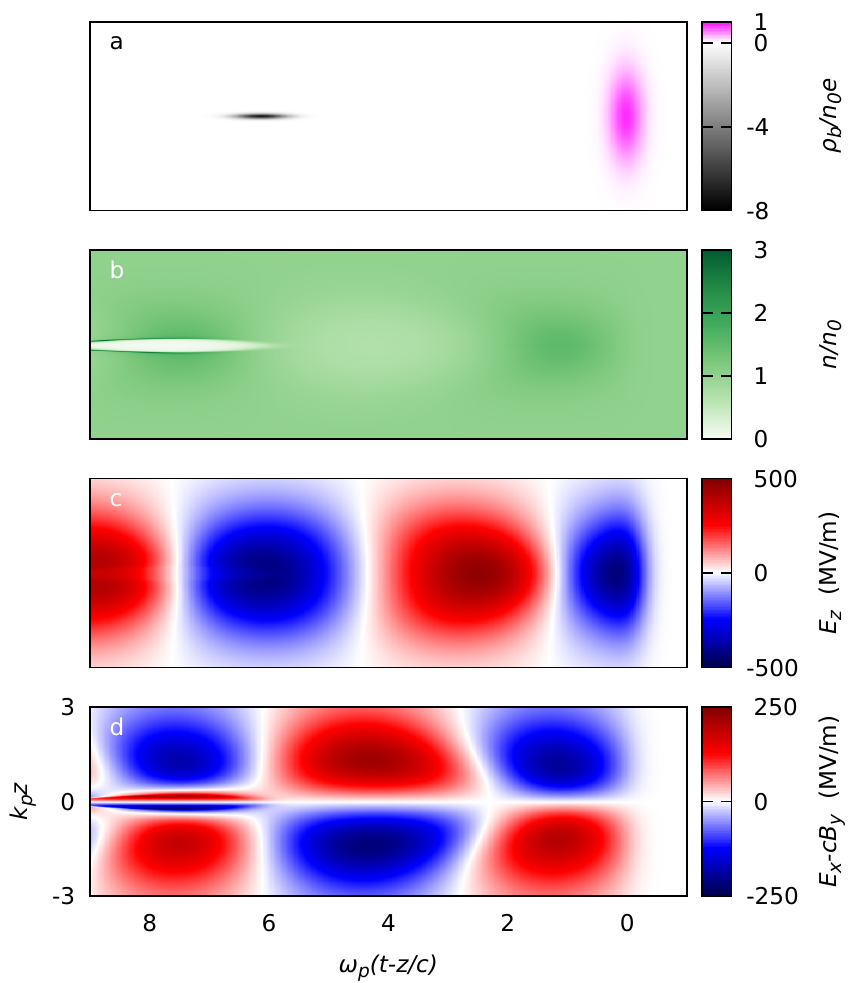}
  \caption{\label{Fig:toy}Toy model configuration used for simulations.  a) A short, positively charged driver (violet) is trailed by an electron witness bunch (black).  b) The driver excites a quasilinear plasma density perturbation, which results in periodic c) longitunal and d) transverse fields.  The plasma also responds to the witness bunch charge, leading to generation of b) a plasma blowout, c) loading of the longitudinal field and d) a modification of the transverse fields.}
\end{figure}

AWAKE Run 1, which gathered data between 2016 and 2018, injected the electron witness bunch off-axis relative to the un-modulated proton beam. A self-selected portion of the witness bunch was trapped and accelerated \cite{pwfa-AWAKE-acceleration-cropped}.  Simulations suggest that the varying wakefield phase prevents effective acceleration during the growth period of the self-modulation process \cite{pwfa-turner-growth-cropped,pwfa-moshuering-AWAKE3d}, which also reduces the charge capture.  In order to better control the acceleration process, AWAKE Run~2 will make use of a separate self-modulation stage, with a short witness bunch injected on-axis into a subsequent acceleration stage, together with the modulated proton beam \cite{pwfa-AWAKE-physics,pwfa-AWAKE-Run2I_PAC21,pwfa-AWAKE-symmetry}. Simulations show that the use of a plasma density step during the self-modulation stage will result in near-constant wakefields during the acceleration process \cite{pwfa-lotov-step,pwfa-lotov-estep}.

The train of driver microbunches have a periodicity determined by the plasma response, and so their collective wakes sum near coherently.  The witness bunch experiences this combined wake from the preceding microbunches.  Full simulations of the microbunch train are computationally expensive due to its length, with a single bunch from the SPS spanning many plasma periods.  However, this overhead can be greatly reduced by exploiting the coherent response to the modulated beam - instead of a train of microbunches, a single, higher-density microbunch is used.  The wake behind this drive bunch mimics that generated by the full bunch train in the second plasma cell.

This ``toy model'' is shown in Fig.~\ref{Fig:toy}.  Lengths are normalised to the plasma wavenumber $k_p=\omega_p/c$, where $\omega_p$ is the plasma frequency and $c$ the vacuum speed of light.  For the baseline plasma density of $n_0=\num{7e14}$~\si{\centi\metre^{-3}}, the plasma electron skin depth $1/k_p=200$~\si{\micro\metre}.  The drive bunch has a length of 40~\si{\micro\metre}, 3000 times shorter than the typical SPS bunch, greatly reducing the computational overhead.

Other parameters of the drive bunch are chosen to reproduce the wake from the full bunch train.  In the experiment, the use of a plasma density step in the modulation stage will result in a plasma wake which is essentially constant during the acceleration stage \cite{pwfa-lotov-estep}. The toy model therefore uses a non-evolving drive bunch with a Lorentz factor of 426, chosen to match the 400~\si{\giga\electronvolt} proton beam delivered by the SPS at CERN.  In order to reproduce the radial extent of the of the wakefield, the drive bunch radius is chosen as 200~\si{\micro\metre} to match the SPS beam.  Finally, the bunch charge is 2.34~\si{\nano\coulomb}, giving a peak beam density of $n_b=0.83 n_0$.  This is significantly higher than the proton beam density used in the experiment, resulting in an unloaded wakefield amplitude of $\sim470$~\si{\mega\volt/\metre}, equivalent to the wake excitation after many microbunches.

This model is exactly that first used by Olsen~\etal\cite{pwfa-olsen-inject}.  Although that work was based on an early concept for the experiment, the wakefield amplitude is close to the value predicted in recent simulations including a plasma density step during the self-modulation stage and a 1~\si{\metre} gap between the self-modulation and acceleration stages \cite{pwfa-AWAKE-symmetry}.  The impact of a higher-amplitude wakefield is considered 
later (see Fig.~\ref{Fig:ropt_2E}).

The witness bunch trails the driver and is accelerated by the wakefield.  The plasma also reacts to the witness bunch charge, which can be beneficial.  Beamloading is the reduction of the accelerating field caused by the witness bunch's own wake, as can be seen in Fig.~\ref{Fig:toy}c.  This effect can be used to locally flatten the accelerating field to reduce the final energy spread of the accelerated bunch.  The witness bunch will also experience additional focussing as the plasma electrons move to compensate its charge.  For a sufficiently dense electron witness bunch, this effect will saturate with the generation of a plasma blowout, or ``bubble'', as seen in Fig.~\ref{Fig:toy}b, exposing the witness bunch to the pure ion column, unshielded by plasma electrons.  This results in a focussing field which increases linearly with transverse displacement, allowing emittance preservation.  As seen in Fig.~\ref{Fig:toy}d, the resulting perturbation to the periodic transverse fields excited by the driver also allows the witness bunch to occupy the region near the peak accelerating field, which corresponds to the point where the unperturbed driver wakefield would switch from focussing to defocussing. The radius of the witness bunch can be chosen according to Eq.~(\ref{Eq:match}), such that the emittance pressure exactly compensates the focussing field in the blowout region.  

\begin{figure}
  \includegraphics{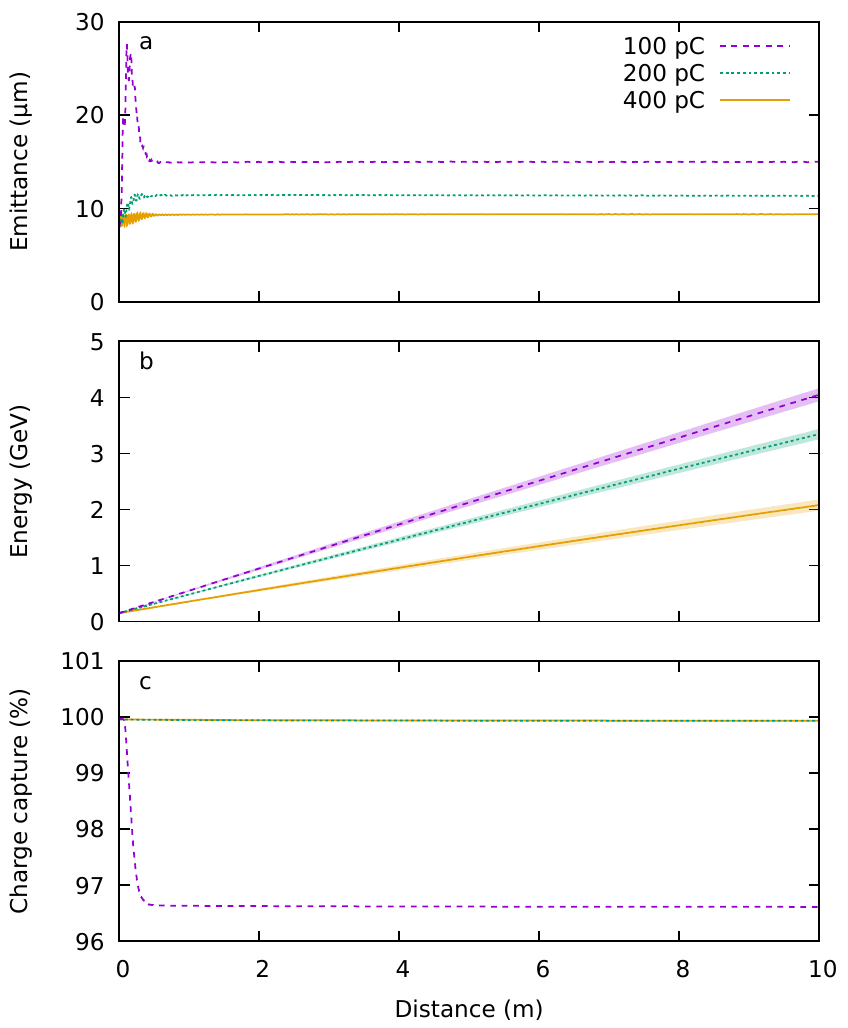}
  \caption{\label{Fig:evolve}Evolution of an injected witness bunch for different initial charges over 10~\si{\metre} acceleration.  Delay in each case is chosen to reduce the relative energy spread while maintaining the charge capture.  a) Emittance, b) energy and c) charge capture.  Shaded areas in (b) show the RMS energy spread.}
\end{figure}

Simulations were carried out using the radially symmetric quasistatic particle-in-cell code LCODE\cite{pic-lotov-lcode,pic-lcode-manual}, using 100,000 equally weighted macroparticles for the witness bunch.  Figure~\ref{Fig:evolve} shows the evolution of the witness bunch for three different witness bunch charges.  In each case, the witness bunch has a Gaussian-ellipsoid profile with a length of 60~\si{\micro\metre}, an initial energy of 150~\si{\mega\electronvolt}, and a normalised emittance of 8~\si{\micro\metre}.  The initial radius is chosen as the matched value of $\sigma_\mathrm{ic}=11.5$~\si{\micro\metre}.  These values differ from those used in \cite{pwfa-olsen-inject} due to the evolution of the experimental design.  Notably, the witness bunch emittance, and so the radius, is significantly higher - in Run~2c, the witness bunch will now be injected through the beam dump for the counterpropagating ionizing laser \cite{pwfa-AWAKE-symmetry}, which will result in stochastic scattering of the bunch prior to injection, increasing the emittance from the 2~\si{\micro\metre} used in \cite{pwfa-olsen-inject}.  It should be noted that the scale length for a plasma accelerator is the plasma electron skin depth, and so the results of this study remain relevant for other configurations with a lower emittance and a higher plasma density.  For example, the scaled length of the witness bunch is five times longer than that used in \cite{pwfa-romeo-assistedmatching} for the \num{1e16}~\si{\centi\metre^{-3}} working point.

As can be seen in Fig.~\ref{Fig:evolve}a, the different initial charges evolve in a similar way, with an initial period of emittance growth before reaching a constant value.  This occurs because the plasma blowout takes some finite time to develop after the start of the bunch, and so the witness bunch head is not matched to the lower focussing fields of the quasilinear plasma wake \cite{pwfa-olsen-inject}.  This causes the radius of the front of the beam to expand and then oscillate, until phase mixing leads to a steady-state solution.  The emittance growth is larger for lower charge as the blowout takes longer to form, resulting in a smaller fraction of the beam exposed to the focussing fields of the ion column.  For the lowest charge of 100~\si{\pico\coulomb}, some of the initial charge is lost, as shown in Fig.~\ref{Fig:evolve}c.  This occurs as the blowout does not form quickly enough to compensate the reduction in the focussing field of the driver wake near the peak accelerating field.  This charge loss is accompanied by a period of decreasing emittance, as electrons with high transverse momentum escape the simulation domain.  After this rapid evolution in the first $\sim50$~\si{\centi\metre} of propagation, the emittance remains essentially constant for the rest of the 10~m acceleration length.  The energy gain, shown in Fig.~\ref{Fig:evolve}b, is essentially linear, with some slight deviation due to dephasing between the witness and the non-evolving driver.  A higher witness bunch charge results in more beamloading, and so a lower acceleration gradient \cite{pwfa-farmer-ipac}.  

\begin{figure}
  \includegraphics{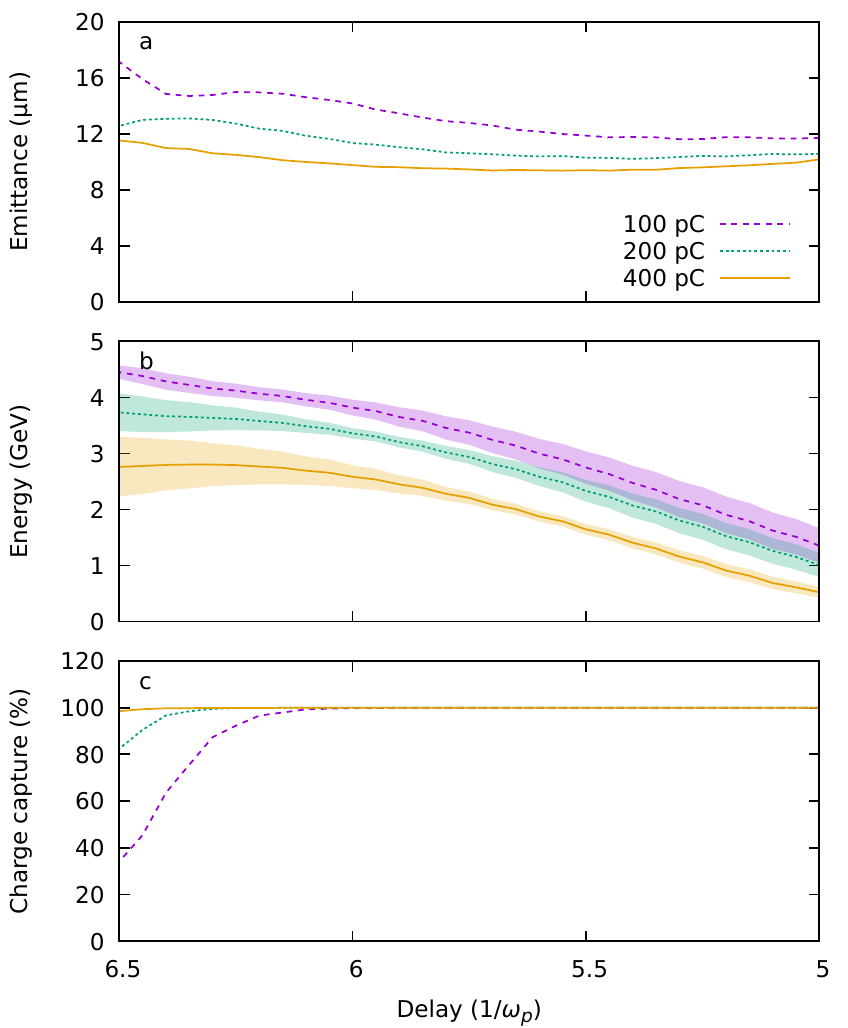}
  \caption{\label{Fig:delayscan_em}Witness bunch parameters after 10~\si{\metre} acceleration for different initial witness charges.  Plots show the influence of the initial witness bunch delay on the final a) emittance, b) energy and c) charge capture.  Shaded areas in (b) show the RMS energy spread.}
\end{figure}

Such simulations can be used as the basis for a parameter scan in order to identify trends and optimal parameters.  These can also serve to identify the tolerances for these parameters, an important consideration for the development of an experiment.  Figure~\ref{Fig:delayscan_em} shows the witness bunch parameters after 10~\si{\metre} of acceleration for varying initial delay between the driver and witness bunches.  Figure~\ref{Fig:delayscan_em}a shows that the emittance depends only weakly on the delay, while the energy gain and final energy spread, shown in Fig.~\ref{Fig:delayscan_em}b, are strongly impacted.  The latter arise as the witness bunch must be carefully placed in order for beamloading to flatten the accelerating field, with different optimal delays for different bunch charges  \cite{pwfa-farmer-ipac}.  The relative energy spread is generally lower for 200~\si{\pico\coulomb} than for 100~\si{\pico\coulomb}, with the latter only achieving a lower energy spread when some of the charge is lost, shown in Fig.~\ref{Fig:delayscan_em}c.  This is because the 200~\si{\pico\coulomb} beam is better able to flatten the accelerating field acting upon it.  Conversely, the 400~\si{\pico\coulomb} witness bunch overloads the wakefield, increasing the relative energy spread.  This beamloading could be further optimised through the use of a witness bunch with a tailored current profile \cite{pwfa-meer-beamloading}, but such optimisations are not a priority for AWAKE Run~2.  Charge capture is only reduced for larger delays, where the witness bunch head can be defocussed before the formation of the blowout.

\section{Figure of Merit for AWAKE}\label{fom}

While parameter scans such as those in Fig.~\ref{Fig:delayscan_em} are certainly informative, they do not provide a definitive measure for the beam quality. Higher witness bunch charge leads to a lower emittance, as discussed above, but can also result in an increase in the relative energy spread.  In some cases, a reduction in energy spread comes at the cost of lower charge capture.

\hl{Furthermore, we expect the slice emittance to vary along the length of the witness bunch, and so considering only the projected emittance of the entire bunch may not give a meaningful measure of the overall beam quality.  One approach is to consider the fraction of the witness bunch for which the transverse emittance is conserved {\cite{pwfa-olsen-inject}}.  However, even in the beam head where the slice emittance increases, a lower emittance could be preserved through beam cleaning, removing some fraction of the charge with large betatron amplitudes.}

Obviously, the relative importance of these various parameters will depend on the application.  There are several potential applications for AWAKE \cite{pwfa-AWAKE-symmetry}, with one potential medium-term application being an electron--solid-target experiment \cite{pwfa-caldwell-particleforAWAKE}.  However, such configurations typically make use of a centimetre-scale target \cite{particle-aubert-beamsize,particle-NA64-beamsize}, which does not place demanding constraints on the beam quality.  It is therefore useful to consider a longer-term potential application, which can serve as a basis for optimisation.  In this study, we will consider the case of an electron--proton collider \cite{pwfa-caldwell-vhep,pwfa-caldwell-particleforAWAKE}, as it fully exploits AWAKE's unique position at CERN.  Here, the figure of merit is the luminosity, which is maximised by increasing the total charge which can be focussed into the cross-section of a counterpropagating proton beam, with the latter having an expected radius on the order of 10~\si{\micro\metre}.  While emittance certainly plays an important role in how easily the accelerated witness bunch can be focussed, chromatic effects due to energy spread can easily dominate.

\begin{figure}
  \includegraphics[width=0.9\columnwidth]{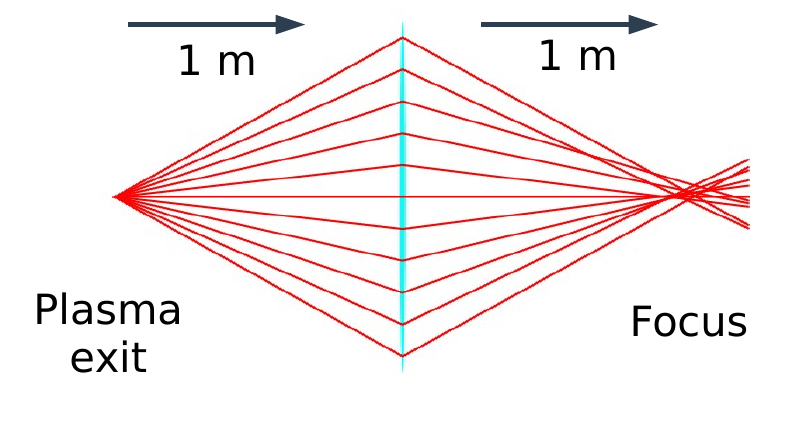}
  \caption{\label{Fig:lens}Schematic for the idealised beam transport model used in the figure of merit.  An ideal focussing element is placed 1~\si{\metre} after the plasma exit. The attainable charge on a 10~\si{\micro\metre} target placed 1~\si{\metre} from the focussing element is measured.}
\end{figure}

We therefore again propose a ``toy model'', this time for a simplified beam transport system, which will serve as our figure of merit for the accelerated witness bunch.  The model is shown in Fig.~\ref{Fig:lens}. The beam is focussed by a single ideal focussing element placed 1~\si{\metre} after the plasma exit.  The total charge that can be focussed within a radius 10~\si{\micro\metre} a further 1~\si{\metre} from the focussing element is then measured.

For the strongly relativistic accelerated witness bunch, space-charge effects can be ignored, and so the propagation of the beam through the toy-model focussing system can readily be calculated on a particle-by-particle basis.  In Cartesian coordinates, the transverse position of a particle at the target location, $x_t$, is:
\begin{equation}\label{proptotarget}
  x_t=x+\frac{2p_xL}{p_z}-\frac{k_fL}{p_z}\left(x+\frac{p_xL}{p_z}\right),
\end{equation}
Here $x$ is the transverse position of the particle, $p_x$ and $p_z$ the transverse and longitudinal momenta, all evaluated at the plasma exit.  $k_f$ is the focussing strength of the ideal focussing element, and $L=1$~\si{\metre} the distance from the plasma exit to the focussing element, and from the focussing element to the target.  Performing the same calculation for $y_t$, the figure of merit can then be written by summing the following inequality over all particles:
\begin{equation}
  C=\sum q \left[x_t^2+y_t^2 < R^2\right]
\end{equation}
where $q$ is the particle charge and
$R=10$~\si{\micro\metre} is the target radius.  $C$ therefore represents the charge-on-target, i.e.~the charge which would impinge on a circular target with a radius of $10$~\si{\micro\metre} after the focussing system shown in Fig.~\ref{Fig:lens}.

Of course, this model is far from complete - any real beam transport system will certainly be more complex, with a longer transport length and higher-order optical elements to introduce corrections to the beam.  However, using a more detailed model may skew the results for the figure of merit, which would essentially lead to designing the accelerator for the transport system, rather than designing the transport system for the accelerator.  The figure of merit does not include the energy of the accelerated bunch, as unlike an increase in energy spread or emittance, this can more readily be compensated, for example by extending the length of acceleration stage.  It should also be noted that any real electron--proton collider would also involve a longer acceleration stage, which would also impact on the beam transport.

However, the great advantage of this figure of merit is that it is application-oriented and provides a single metric to characterise the accelerated beam.  Although the values for $C$ will be subject to change as the design progresses, trends are expected to persist, and so it provides a basis for the optimisation of the acceleration process, and a cost--reward basis for future design decisions.  A further benefit is that this figure of merit is not sensitive to outliers in the phase-space distribution, unlike the statistical measures used in section~\ref{simconf}.  Outliers certainly dominate for the 100~\si{\pico\coulomb} case shown in Fig.~\ref{Fig:evolve}, where the emittance decreases significantly as particles leave the simulation domain.  While this can be avoided by making use of more complex statistical techniques, this does rely on somewhat arbitrary choices. The toy-model transport system avoids these statistical limitations by making use of the entire beam distribution.

\begin{figure}
  \includegraphics{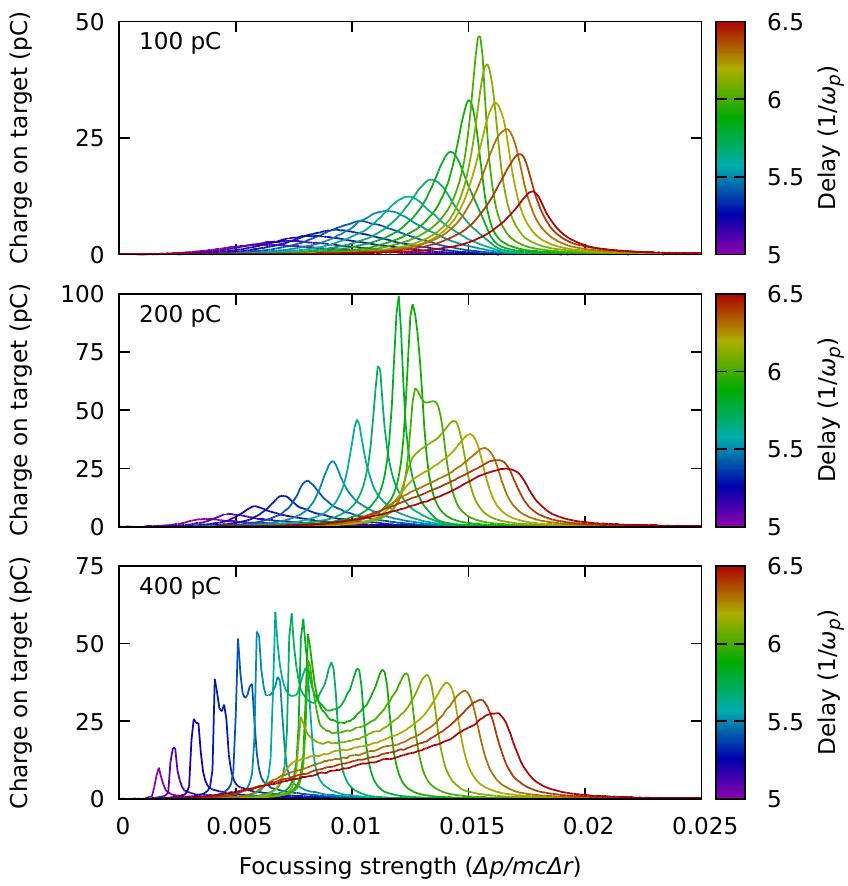}
  \caption{\label{Fig:focus}Results from the toy model beam transport shown in Fig.~\ref{Fig:lens}: the charge-on-target $C$ as a function of the focussing strength.  Each line represents a single simulation, colour coded according to the witness bunch delay at injection.  Each plot shows a different initial witness charge.}
\end{figure}

This model is used in Fig.~\ref{Fig:focus} to determine the charge-on-target as a function of the focussing force applied to the beam.  Different delays are shown, with each plot showing a different initial witness bunch charge.  \hl{Of the three charges considered, the 200-{\si{\pico\coulomb}} witness bunch delivers the most charge-on-target.  Despite the 400~{\si{\pico\coulomb}} witness bunch having a lower emittance at the plasma exit, as shown in Fig.~{\ref{Fig:delayscan_em}}, it delivers significantly less charge-on-target than the 200-{\si{\pico\coulomb}} witness bunch.  This is due to the larger relative energy spread, which increases chromatic effects in the focussing system and significantly reduces the fraction of the charge which can be transported to the 10-{\si{\micro\metre}} target.  Higher charge reduces the energy gain via beamloading, and so a lower focussing field is required.  For the 400~{\si{\pico\coulomb}} case, the curves develop complex structure.  In this case, a single bunch with two distinct foci is indicative of an energy spectrum with two peaks.}

\begin{figure}
  \includegraphics{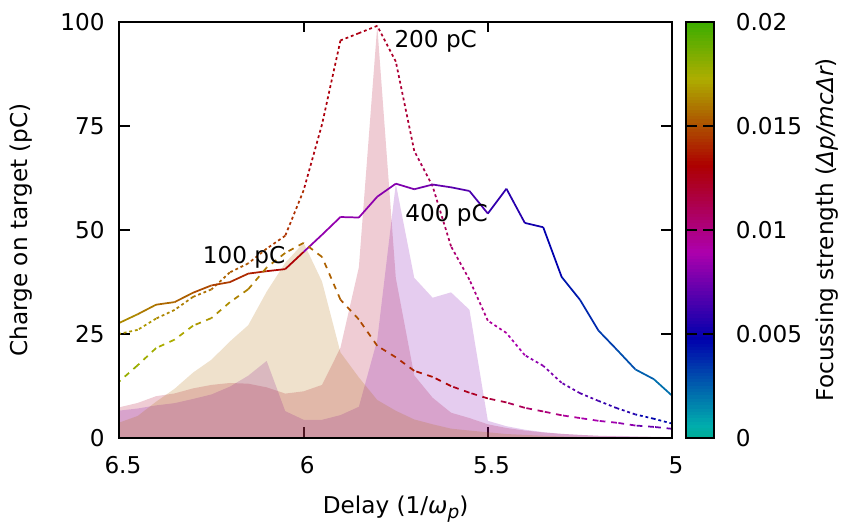}
  \caption{\label{Fig:opt}The results of Fig.~\ref{Fig:focus} can be summarised by the maximum attainable charge-on-target for each delay.  Lines show adaptive focussing, relevant for the case of a constant delay.  The filled curves show fixed focussing (chosen to give the maximum charge-on-target for the optimal delay), relevant for the case of jitter.}
\end{figure}

Although elegant, the plots shown in Fig.~{\ref{Fig:focus}} do not lend themselves to comparisons between many different parameter sets.  \hl{They can be summarised by considering the maximum achievable charge-on-target for a given delay, as shown in Fig.~{\ref{Fig:opt}} (note the x and colourbar axes have been swapped relative to Fig.~{\ref{Fig:focus}}).  Lines show the result of optimising the focussing strength for each delay in order to maximize the charge-on-target.  This corresponds to taking the peak of each of the lines in Fig.~{\ref{Fig:focus}}.  Physically, this corresponds to adaptive focussing - a shift in the delay can be partially compensated by changing the focussing strength during the beam transport.  This therefore gives a measure for the required tunability of the delay.}

\hl{However, in an experimental setting, beam parameters may fluctuate on a shot-to-shot basis.  In this case, the focussing optics cannot be optimised for every shot.  The filled curves in Fig.~{\ref{Fig:opt}} illustrate the impact varying the delay for fixed focussing, i.e.\ taking the value of each line in Fig.~{\ref{Fig:focus}} at the position of the global peak.  These filled curves therefore illustrate the impact of jitter in the bunch delay, and so give a measure of the of the required stability of the delay.}

As can be seen, the acceleration scheme is much more robust to a constant offset from the ideal delay than it is to jitter.  This plot therefore allows the tolerances for both tunability and stability to be gauged, and can readily be applied to other parameters.  For the 100~{\si{\pico\coulomb}} case, the full-width for adaptive focussing at 80\% of the peak achievable charge-on-target is $\sim0.21/\omega_p$, corresponding to a required tunability (half-width) of $\pm$70~{\si{\femto\second}}.  For the case of fixed-focussing, the full-width is $\sim0.13/\omega_p$ corresponding to a required stability (half-width) of $\pm$45{\si{\femto\second}}.

This metric is also robust against numerical fluctuations introduced via simulation methods.  The data points used to create the parameter scans in Figs.~\ref{Fig:delayscan_em}, \ref{Fig:focus} and \ref{Fig:opt} show the result of single simulations.  Although the random-number generator used to create the particle bunch introduces some degree of stochastic behaviour, repeat runs show that the fluctuation of the charge-on-target is $\ll1$~\si{\pico\coulomb} for the 100~\si{\pico\coulomb} witness bunch, and so taking average values over multiple simulation runs was not considered necessary.

\section{Self-matching of the plasma wakefields to the witness bunch}\label{match}
Having discussed the need for a figure of merit and demonstrated its utility for parameter scans, we may now use it to investigate the tolerances for experimental parameters of interest.

\begin{figure}
  \includegraphics{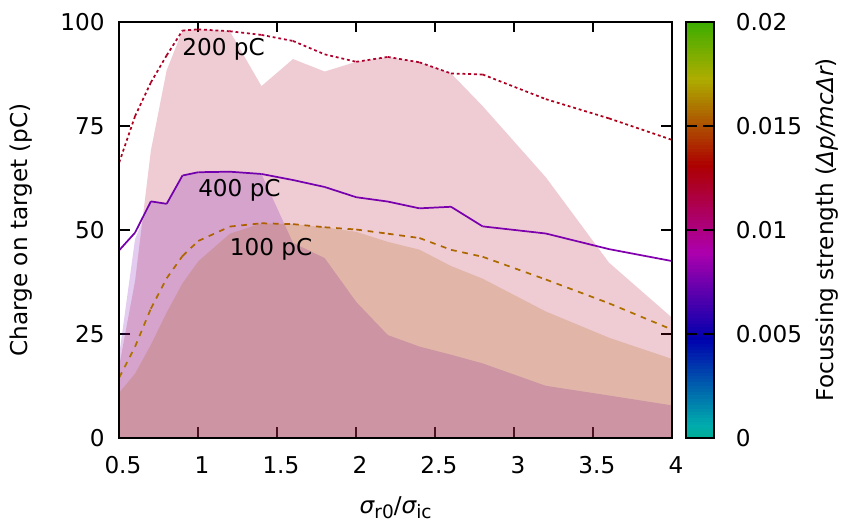}
  \caption{\label{Fig:ropt}a) The fraction of charge-on-target, $C$, for different initial witness bunch radii.  In each case, the delay is chosen as the optimal for $r_0/\sigma_\mathrm{ic}=1$.  As in Fig.~\ref{Fig:opt}, lines show adaptive focussing, filled curves show fixed focussing.}
\end{figure}


First among these is transverse size of the witness bunch.  For the AWAKE experiment, achieving the matched radius for the ion channel is challenging {\cite{pwfa-ramjiawan-AWAKEoptimizer}} due to the constraints of the electron beamline.  For a laser-driven wakefield accelerator, higher plasma densities are typically used, and so achieving the matched radius may be even more challenging.  Figure~\ref{Fig:ropt} shows the influence of varying the initial witness bunch radius relative to the ion-channel matched value for different witness charges.  The delay relative to the witness bunch is fixed according to the results of Fig.~\ref{Fig:opt}, as the longitudinal field depends predominantly on the witness bunch current, rather than charge density.  It is immediately apparent that the tolerances are very large, with the charge-on-target decreasing by only a few percent if the radius is doubled.

This broad tolerance is due to the plasma response to the witness bunch.  The focussing fields increase along the length of the witness bunch {\cite{pwfa-farmer-ipac}}, and so the matched radius for the head will be larger than that for the tail.  Increasing the witness bunch radius therefore results in better matching of the head.  However, this increase in the bunch radius will reduce the charge density, causing the focussing fields due to the plasma response to increase more slowly along the bunch length.  In this way, a larger-radius witness bunch experiences a lower average focussing field, which in turn increases the optimal radius.  It is this ``self-matching'' of the plasma wakefield to the bunch radius which gives rise to these broad tolerances.

This is borne out by the fact that the optimal radius for the 100~\si{\pico\coulomb} case is larger than the matched radius for the ion channel.  Since the charge density is lower, the blowout takes longer to form, and less of the bunch will sit in the blowout region. 

This behaviour is unique to injection into a quasilinear wake - for acceleration in a blowout, the entire witness bunch feels the focussing fields of the ion channel and the matched radius is essentially independent of the witness bunch radius.

\section{Additional witness bunch tolerance studies}\label{everythingelse}

\begin{figure}
  \includegraphics{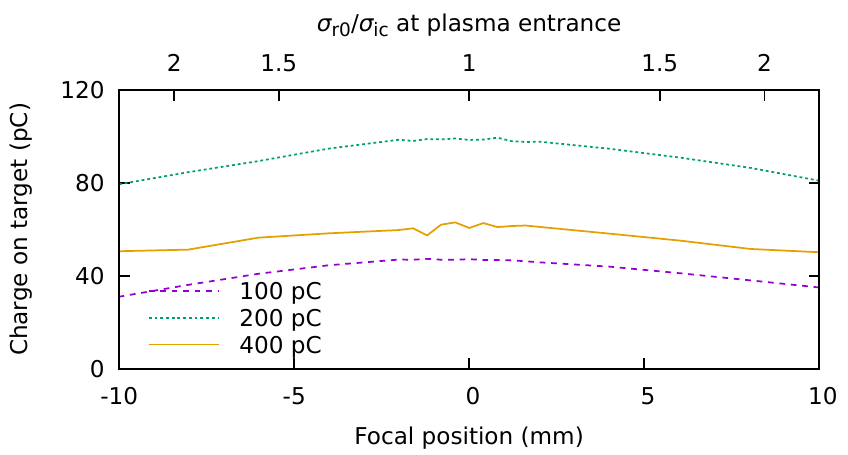}
  \caption{\label{Fig:movef}Influence of shifting the focal position for a 100~\si{\pico\coulomb} beam with 8~\si{\micro\metre} normalised emittance.  The waist size is chosen as the matched radius for the ion column, with a negative focal position corresponding to the focus placed before the plasma entrance.}
\end{figure}

A closely related parameter of interest is the focal position of the witness bunch relative to the plasma entrance, shown in Fig.~\ref{Fig:movef}.  Although the tolerances here are also large, it is notable that shifting the focus decreases the charge-on-target more rapidly than the equivalent increase in the beam radius if the focus is kept at the plasma entrance.  It is also interesting to note that the increase in charge-on-target achieved by increasing the radius for the 100~\si{\pico\coulomb} case is not replicated by shifting the focus.

\begin{figure}
  \includegraphics{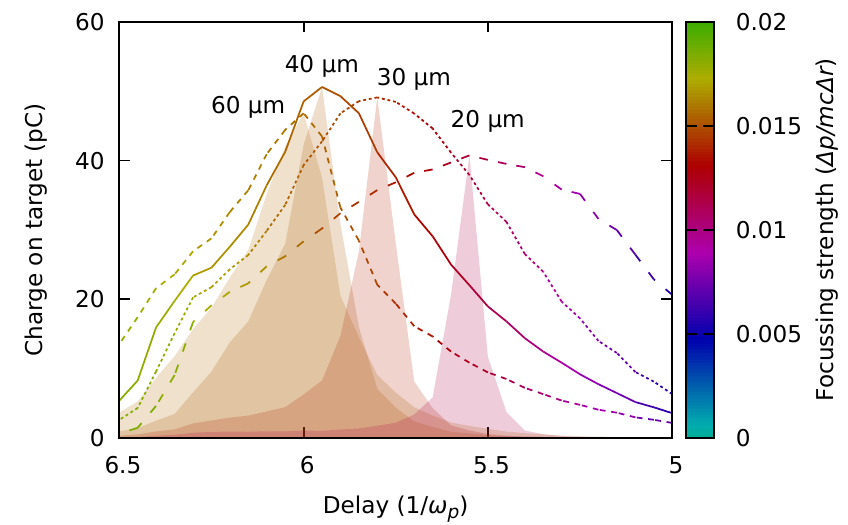}
  \caption{\label{Fig:opt_l}Influence of the duration of a 100-\si{\pico\coulomb} witness bunch on the achievable charge-on-target and timing tolerances.  The bunch radius is chosen as the matched radius for the ion column.}
\end{figure}

\begin{figure}
  \includegraphics{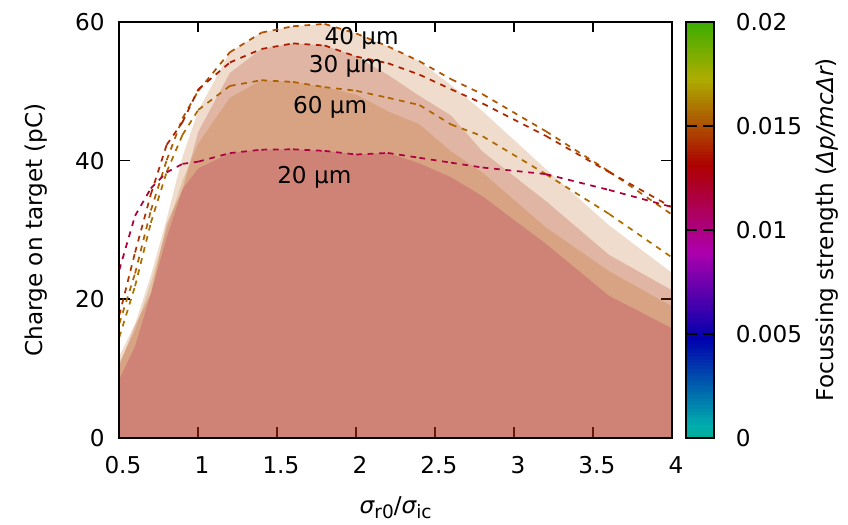}
  \caption{\label{Fig:ropt_l}Influence of the duration of a 100-\si{\pico\coulomb} witness bunch on the achievable charge-on-target and radial matching tolerances.  The delay in each case is chosen as the optimal for $\sigma_0=\sigma_\mathrm{ic}=1$.}
\end{figure}

Simulations carried out for varying witness bunch duration show that this has only a weak influence on the achievable charge-on-target, as seen in Fig.~\ref{Fig:opt_l} for a bunch charge of 100 \si{\pico\coulomb} and a radius matched to the ion column.  When keeping the bunch charge constant, the blowout forms more rapidly for a shorter bunch, but the fraction of the bunch within the blowout, and so the transverse matching, is not strongly affected.  A shorter witness does give a low energy spread over a wider range of delays, as seen from the wide tuning tolerance for the 20~\si{\micro\metre} duration.  However, this narrow energy spread also increases the sensitivity to timing jitter, as the fraction of particles with a given energy drops more rapidly as the mean energy shifts away.

\begin{figure}
  \includegraphics{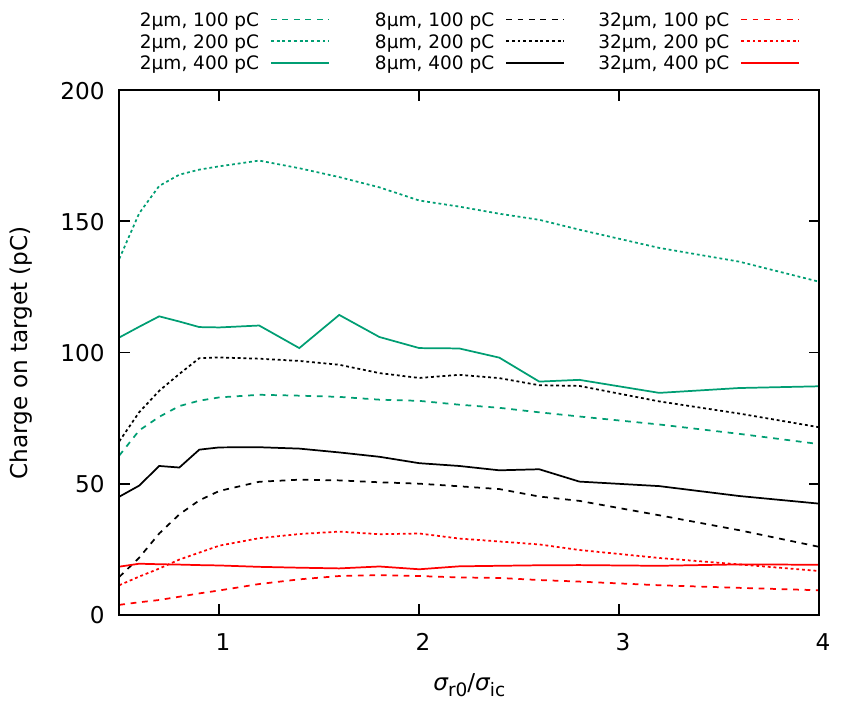}
  \caption{\label{Fig:rem}Charge-on-target for different initial charges and emittances.  The delay for each case is chosen as the optimal for $r_0/\sigma_\mathrm{ic}=1$.}
\end{figure}

As discussed in section~\ref{simconf}, for the planned AWAKE Run 2 configuration, the witness bunch emittance at injection is dominated by scattering as it passes through the laser beam dump.  Alternative designs could reduce, or indeed increase, this contribution.  Figure~\ref{Fig:rem} shows the influence of varying both the charge and initial emittance of the witness bunch.  For larger emittances, the optimal radius is increased relative to the ion-channel matched value, as the reduction in the witness charge density results in less of the beam sitting in the blowout region.  Although self-matching of the wakefields results in a broad tolerance for the initial radius in all cases, the absolute beam quality is higher for a lower initial emittance.  This is because self-matching allows the initial emittance to be conserved, and so a smaller initial emittance results in a smaller final emittance, which is more easily focussed onto the target.

From a design perspective, it should be noted that although the acceleration process is only weakly influenced by the witness bunch radius at the injection point, any variation will also change the bunch radius at the point where it passes through the laser beam dump, changing the contribution due to scattering.  This in turn will further modify the bunch radius at the injection point, as well as the emittance and the focal position.  A combined treatment of these effects has been carried out during the beamline design using numerical optimizers \cite{pwfa-ramjiawan-AWAKEoptimizer}.



\begin{figure}
  \includegraphics{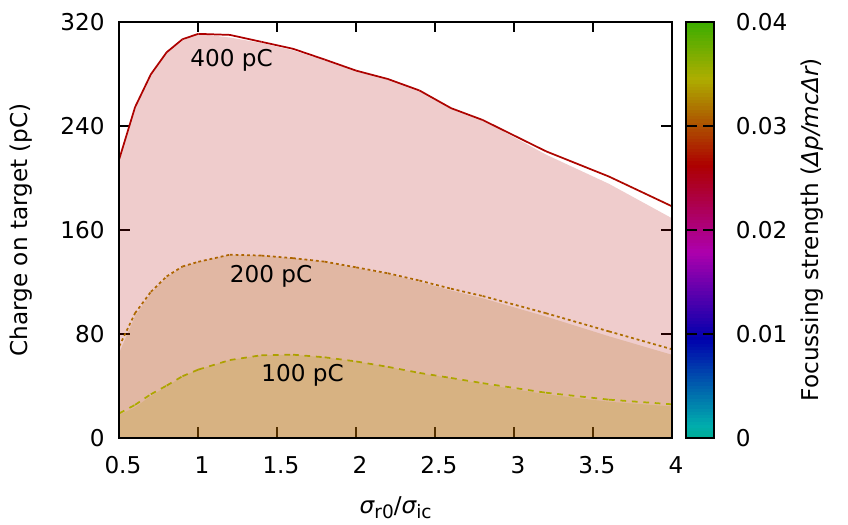}
  \caption{\label{Fig:ropt_2E}As Fig.~\ref{Fig:ropt}, but with double the charge in the drive beam.}
\end{figure}


The optimal parameters for the witness bunch will also depend on the wakefield excited by the drive beam.  Figure~\ref{Fig:ropt_2E} shows the result of scanning the witness bunch radius at the plasma entrance when the charge of the drive beam is increased by a factor two, giving a peak wakefield of $\sim950$~\si{\mega\electronvolt/\metre} (slightly more than double due to increasing nonlinearity).  This is similar to the wakefields achievable if the gap between the self-modulation and acceleration stages can be avoided \cite{pwfa-lotov-estep,pwfa-AWAKE-symmetry}.  In this case, the 400~\si{\pico\coulomb} beam gives best results.  Again, these results are dominated by the energy spread, with the higher-charge witness bunch better suited to load the higher-amplitude wakefields.  Note that the required focussing strength is higher, due to the increased energy gain.  The fraction of the initial beam achievable on target is also higher, likely as the stronger wakefield reduces the expansion of the witness bunch head prior to the formation of the blowout, combined with the faster blowout formation for higher charge.  The tolerances for the witness timing (not shown) are similar to those observed for the lower-amplitude wakefield.

One parameter not considered in this work is the transverse alignment of the witness bunch relative to the wakefield axis.  Such studies would require the use of fully-3D PIC simulations rather than the radially symmetric model used here.  Misalignment causes the witness to spread transversely as particles undergo betatron oscillations in the wakefield excited by the drive beam.  For the Run~2c beamline, the shot-to-shot jitter in the witness bunch alignment is small, on the scale of a few microns \cite{pwfa-ramjiawan-AWAKEoptimizer}, which does not significantly modify the quality of the accelerated beam \cite{pwfa-olsen-inject}.  The jitter in the proton beam position is significantly larger, and its influence on the acceleration process will be investigated in a separate study.  However, any collider would necessarily require that the stability of the proton beam be addressed, and so the assumption here of aligned beams is reasonable.

\section{Conclusion}\label{conc}
\hl{Acceleration in a quasilinear plasma wake offers high field gradients, while avoiding self-trapping of the background plasma electrons.  However, in this regime the plasma wakefield also responds to the witness bunch, and so the quality of the accelerated bunch depends on its initial parameters at the injection point.  Furthermore, quantities such as the emittance may vary along the bunch length.

In order to characterise the overall quality of the accelerated witness bunch, in this work we develop a figure of merit, $C$, based on an idealised beam transport system for a potential high-energy application of AWAKE, an electron--proton collider.  This allows different initial parameter sets to be easily compared.  As the design concept is necessarily in an early stage, a toy model is used with a single focussing element.  Although the absolute values for the figure of merit are therefore subject to change, it nonetheless provides a basis for the optimisation of the acceleration process.  As the design progresses, more advanced models could be adopted.  Similarly, alternative models could be developed for different applications, or indeed for alternative acceleration schemes, making this approach widely applicable.  We show this metric naturally gives rise to constraints on the stability and tunability required for high-quality acceleration.

We make use of this figure or merit to study the tolerances at the injection point for a witness bunch injected into a quasilinear wake.  We find the tolerances for the witness bunch radius at the injection point are extremely broad, with the achievable charge-on-target decreasing by only a few percent if the radius is doubled.  This is attributed to ``self-matching'' of the plasma wakefields, in which a larger-radius witness bunch experiences lower focussing fields.  This previously unobserved phenomenon greatly reduces the degradation of beam quality caused by mismatch.

While self-matching results in broad tolerances for the initial radius, a lower initial emittance still provides an improvement in the absolute beam quality.  This is due to self-matching allowing the initial emittance to be preserved even if the matching condition is not exactly satisfied.  It is also shown that, in addition to providing larger acceleration gradients, larger-amplitude wakes also result in a higher overall beam quality, as the matched radius for the quasilinear wake approaches that of the blowout region.}

\section{Acknowledgements}
The authors would like to thank Alexander Gorn and Konstantin Lotov for support with LCODE, and Allen Caldwell and Matthew Wing for discussions on high-energy applications for AWAKE.

\bibliography{bib}

\end{document}